\documentclass[a4paper,11pt]{article}
\usepackage{pos}

\title{$t\bar{t}t\bar{t}$: NLO QCD corrections in production and decays for the $4\ell$ channel}
\author*[a]{Nikolaos Dimitrakopoulos}
\affiliation[a]{Institute for Theoretical Particle Physics and Cosmology, RWTH Aachen University, \\
D-52056 Aachen, Germany}

\emailAdd{ndimitrak@physik.rwth-aachen.de}

\abstract{We briefly summarise the results for the four top-quark production process in the 4$\ell$ decay channel at  NLO accuracy in perturbative QCD. We employ the narrow-width approximation for the treatment of the unstable particles, preserving spin correlations to NLO accuracy throughout our computations.  The NLO QCD corrections are applied to both the production and decays of the four tops.  We investigate the overall size of the NLO QCD corrections and the impact of neglecting higher-order effects in top-quark decays.  \\

\begin{flushright} P3H-24-073,  
TTK-24-41 \end{flushright}}

\FullConference{42nd International Conference on High Energy Physics (ICHEP2024)\\
18-24 July 2024\\
Prague, Czech Republic\\}

\begin{document}
\maketitle

\section{Introduction}
The simultaneous production of four top quarks is one of the rarest processes at the LHC owing to its very small cross section. In spite of that, the four-top production process is both intriguing and important as it provides a direct way to probe the top Yukawa coupling, complementary to $t\bar{t}H$ production. On top of that, this process is highly sensitive to many beyond the Standard Model (SM) scenarios, where extra intermediate heavy resonances can decay to a pair of top quarks, leading to potential deviations from the SM predictions. Despite its very small cross section, in 2023, the ATLAS and CMS collaborations announced the discovery of four-top production \cite{CMS:2023ftu, ATLAS:2023ajo}, by studying the $4\ell, 3\ell$ and $2\ell$ same-sign signal regions. Driven by this exciting discovery, accurate and high-precision theoretical calculations for this process are now more relevant than ever. 

The first theoretical results for the stable four-top production at NLO in QCD were obtained in Ref. \cite{Bevilacqua:2012em} and later also in Ref. \cite{Maltoni:2015ena, Alwall:2014hca}. In Ref. \cite{Frederix:2017wme}, the so-called complete NLO predictions became available for this process, by considering both QCD and EW corrections to all LO contributions. Additionally, results at next-to-leading logarithmic accuracy are available in the literature in Ref. \cite{vanBeekveld:2022hty}, offering the most accurate predictions for the stable four-top production to date. However, in all of the above studies, the decays of the top quarks were not included in the calculations. Incorporating top-quark decays is crucial, as it provides further information on all particles in the final state and allows for a more realistic study of the fiducial phase space. Towards that direction, in Ref. \cite{Jezo:2021smh} the decays of the top quarks in the $1\ell$ + jets channel were modelled with LO accuracy, preserving spin correlations to the same order in the perturbative expansion. In this study, the theoretical predictions were then matched to Parton Shower (PS) using the POWHEG framework. Nevertheless, results for other signal regions were not provided while the effects of QCD corrections to the decay products were only described in the soft-collinear approximation from the PS. In Ref. \cite{Dimitrakopoulos:2024qib} we provided results at NLO in QCD for the four-top production and decays in the $4\ell$ channel using the narrow-width approximation (NWA). By doing so, the emission of hard radiation is well described in both stages and spin correlations are preserved with NLO accuracy throughout our calculations without any approximation. Here, we will report on the main results obtained in this work. In Section \ref{sec2}, we briefly describe the different methods that we used to provide our results in the NWA.
In Section \ref{sec3} we present the results at the integrated fiducial cross-section level, highlighting the importance of incorporating higher-order effects in both the production and decays. Finally, some results at the differential cross-section level are provided in Section \ref{sec4}.

\section{Description of the calculation} \label{sec2}
In Ref. \cite{Dimitrakopoulos:2024qib} the NWA has been employed for the treatment of both the top quarks and the $W$ bosons, allowing us to separate the production and the decay stages. In this approximation, the Breit Wigner propagator simplifies to a delta function in the limit of $\Gamma/m \to 0$, where $\Gamma$ is the width of the unstable particle and $m$ is its mass
\begin{equation}
    \lim_{\Gamma/m \to 0}\dfrac{1}{(p^2-m^2)^2 + m^2\Gamma^2} = \dfrac{\pi}{m\Gamma}\delta(p^2-m^2)\,.
\end{equation}
However, at NLO in QCD several NWA approaches are possible. In the first approach, which we label $\rm NLO_{full}$, the QCD corrections are applied in both the production and decays by utilizing the NLO top-quark width ($\Gamma_t^{\rm NLO}$) everywhere in the calculation. If on the other hand, the $\Gamma_t$ is treated as a perturbative parameter, an expansion in terms of the strong coupling constant can be performed, which will eventually lead to the following equation:
\begin{equation}
    d\sigma_{\rm exp}^{\rm NLO} = d\sigma^{\rm NLO} = d\sigma_{\rm full}^{\rm NLO} \times \left(\dfrac{\Gamma_t^{\rm NLO}}{\Gamma_t^{\rm LO}}\right)^4 - d\sigma^{\rm LO}\times \dfrac{4(\Gamma_t^{\rm NLO} - \Gamma_t^{\rm LO})}{\Gamma_t^{\rm LO}}\,.
\end{equation}
This approach is denoted as $\rm NLO_{exp}$, or simply $\rm NLO$ since it was used as a default one throughout our computations. In this case, some higher-order effects that are included in the $\rm NLO_{full}$ case when top quarks decay with LO accuracy, are properly removed. Finally, we also consider the case where the QCD corrections are solely applied at the production stage of the four top quarks. In this case, we use the LO top-quark width ($\Gamma_t^{\rm LO}$) and refer to this approach as $\rm NLO_{LO_{dec}}$. 

% The numerical values that we used for the top-quark widths are given below:
% %
% \begin{equation}
%     \Gamma_t^{\rm NLO} = 1.3535983 \rm GeV, \quad \quad \quad \Gamma_t^{\rm LO} = 1.4806842 \rm GeV
% \end{equation}
% %

All our results have been obtained with the help of the \textsc{Helac-Nlo} Monte-Carlo framework \cite{Bevilacqua:2011xh} using a center of mass energy of $\sqrt{s} = 13.6 \; \rm TeV$. Trying to be as inclusive as possible we have applied the following cuts to all final states
\begin{equation}
\begin{array}{lll}
 p_{T,\,\ell}>25 ~{\rm GeV}\,,    
 &\quad \quad \quad \quad\quad|y_\ell|<2.5\,,&
\quad \quad \quad \quad \quad
\Delta R_{\ell
 \ell} > 0.4\,,\\[0.2cm]
p_{T,\,b}>25 ~{\rm GeV}\,,  
&\quad \quad\quad\quad\quad |y_b|<2.5 \,, 
 &\quad \quad\quad \quad \quad
\Delta R_{bb}>0.4\,,
\end{array}
\end{equation}
where we have required at least 4 $b$-jets and exactly 4 charged leptons. We also set no restriction on the missing transverse momentum due to the presence of four neutrinos in the final state.

\section{Integrated fiducial cross sections} \label{sec3}
In all of our calculations, the default setup utilizes the $\rm (N)LO$ MSHT20 PDF set for our $\rm (N)LO$ calculations and a dynamical scale choice for the renormalization ($\mu_R$) and the factorization ($\mu_F$) scale $\mu_0 = \mu_R = \mu_F = E_T/4$, defined as:
\begin{equation}
    E_T = \sum_{i=1,2}\sqrt{m_t^2 + p_T^2(t_i)} + \sum_{i=1,2}\sqrt{m_t^2 + p_T^2(\bar{t}_i)}\,.
\end{equation}
The fiducial cross sections at LO and NLO in QCD are
\begin{equation}
    \sigma^{\rm LO} = 4.7479(3)^{+74\%}_{-40\%} \; \rm ab, \quad \quad \quad \sigma^{\rm NLO} = 5.170(3)^{+12\%}_{-20\%} \; \rm ab \,.
\end{equation}
The scale uncertainties are estimated using the standard $7-$point variation where both $\mu_R$ and $\mu_F$ are varied independently around the central scale choice $\mu_0$ according to
\begin{equation}
    \left( \dfrac{\mu_R}{\mu_0}, \dfrac{\mu_F}{\mu_0}\right) = \left\{ (1,1), (2,1), (0.5,1), (1,2), (1,0.5), (2,2), (0.5,0.5)\right\} \,.
\end{equation}
We have observed a significant reduction in the size of the scale uncertainties when QCD corrections are applied, from $74\%$ at LO to $20\%$ at NLO. In addition to the dynamical scale choice, the fixed scale choice, defined as  $\mu_0 = 2m_t$, has also been studied. Notably, the NLO predictions for the two scale settings differ by up to $5\%$ at the integrated (fiducial) cross-section level and remain consistent within the range of the scale uncertainties. The PDF uncertainties for this process are at the level of $4\%$, making them relatively small compared to the scale uncertainties. Lastly, we have confirmed that the differences between the NLO results obtained with various PDF sets are less than $1\%$.

Another interesting feature worth investigating is the impact of QCD corrections on the top-quark decays as well as the effects of the expansion of $\Gamma_t$ in our calculations. The fiducial cross sections for the relevant NWA cases are given below: 
\begin{equation}
    \sigma_{\rm exp}^{\rm NLO} = 5.170(3)^{+12\%}_{-20\%} \rm \; ab, \quad \quad \sigma_{\rm LO_{dec}}^{\rm NLO} = 5.646(3)^{+22\%}_{-23\%} \; ab, \quad \quad \sigma_{\rm full}^{\rm NLO} = 5.735(3)^{+2\%}_{-15\%} \; ab\,.
\end{equation}
We have noticed that the impact of higher-order QCD effects on the decays is $9.2\%$ when compared to our default setup, while the higher-order effects that are relevant in the $\rm NLO_{full}$ case are about $10.9\%$. Both differences are well within the size of the theoretical uncertainties, making the various NWA treatments consistent within the overall corresponding theoretical errors. In addition, incorporating QCD corrections in both the production and decays reduces the magnitude of the scale uncertainties from $23\%$ to $20\%$. Finally, the size of the uncertainties obtained from scale variation for the $\rm NLO_{full}$ setup is the smallest accounting for $15\%$. 

\section{Differential cross-section distributions} \label{sec4}
\begin{figure}[t!]
        \centering        
        \includegraphics[width=0.48\linewidth]{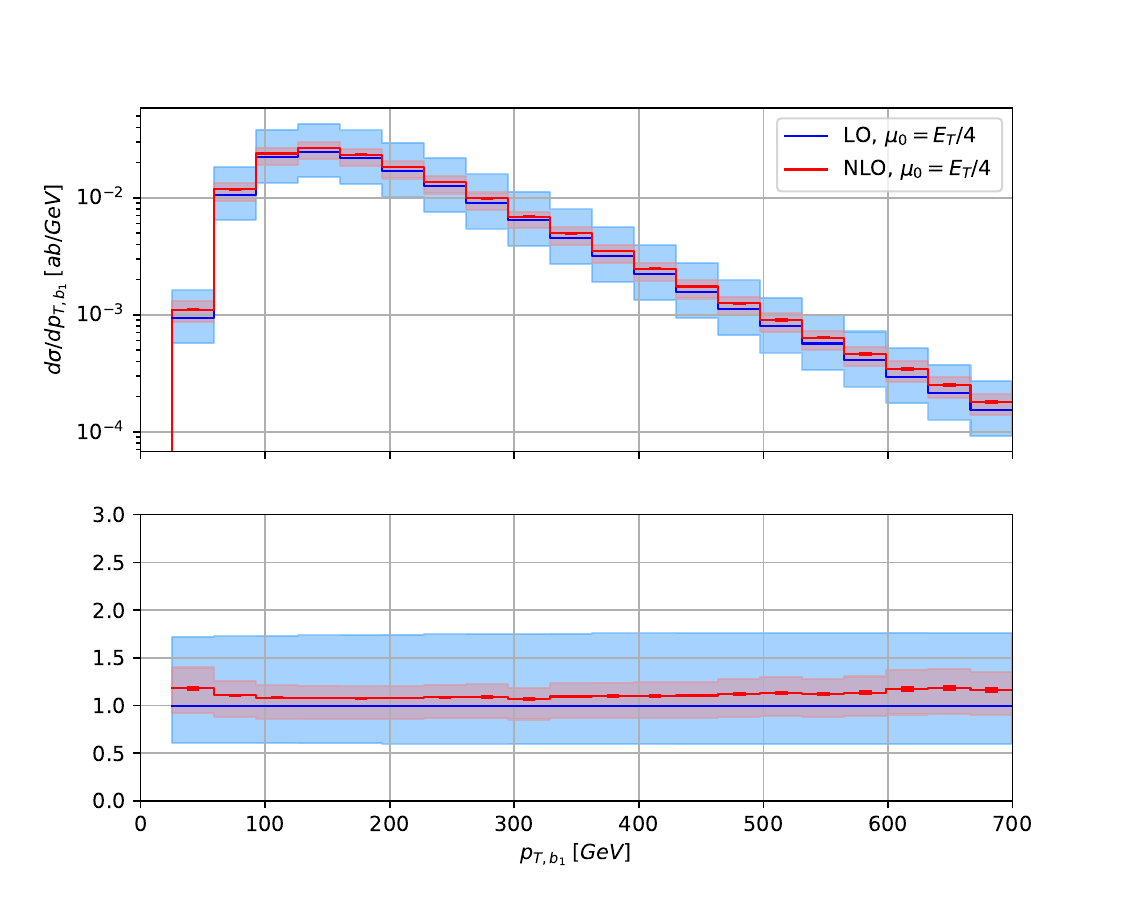}
        \includegraphics[width=0.48\linewidth]{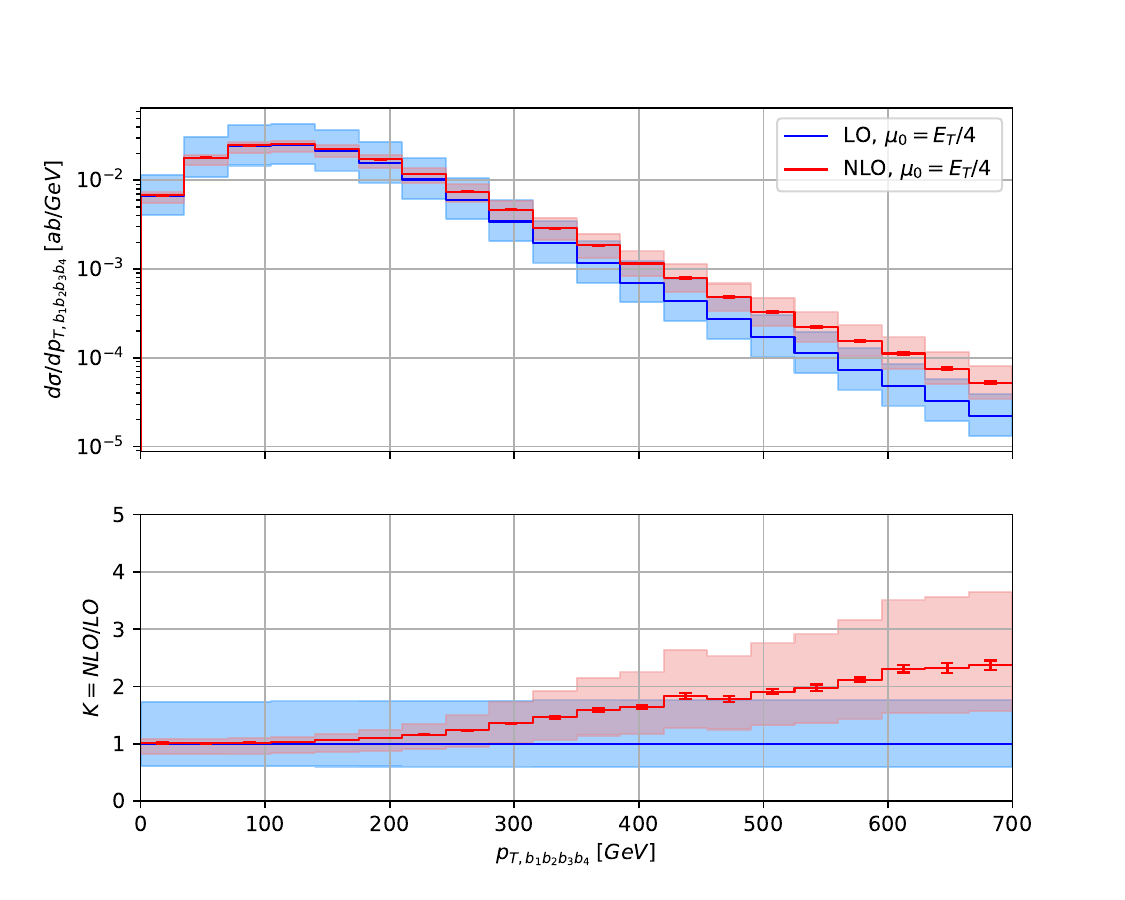}
        \caption{\textit{Differential cross-section distributions for the $pp \to t\bar{t}t\bar{t}$ process in the $4\ell$ channel for the $p_{T,\,b_1}$ (left) and $p_{T,\,b_1b_2b_3b_4}$ (right) observables. In both plots, the upper panels display the absolute LO and NLO predictions with blue and red colors respectively. In the bottom panels we plot the differential $\mathcal{K}$ factor including the scale uncertainties relative to the LO predictions. Figures are taken from Ref. \cite{Dimitrakopoulos:2024qib}.}}
         \label{lonlo.png}
\end{figure}

To estimate the impact of higher-order QCD corrections in different regions of the phase space it is important to present results also at the differential cross-section level. For the majority of the observables that we have studied, the NLO QCD corrections at the differential level are similar in size to those at the integrated cross-section level. Therefore, in these cases, the NLO predictions were covered by the large LO uncertainty bands. This is illustrated in Figure \ref{lonlo.png} (on the left)  where we show the transverse momentum of the hardest $b$-jet. Focusing on the bottom panel where the differential $\mathcal{K}$ factor is depicted, we could observe that the higher-order QCD corrections reach a maximum of $20\%$, with the NLO scale uncertainties being of the same size. Nevertheless, for certain types of observables, QCD corrections can become significantly large in specific regions of the phase space. To demonstrate this, in Figure \ref{lonlo.png} (on the right)  we present the transverse momentum of the system comprising the four hardest $b$-jets. In this case, the NLO QCD corrections become excessively large in the tail of the distribution, where the $\mathcal{K}$ factor reaches values up to $2.2$. In addition, the NLO theoretical uncertainties are also quite high in these phase-space regions, with values as large as $50\%$. These large QCD corrections arise from the presence of a highly energetic light jet emitted at the production stage, which recoils against the system of the four hardest $b$-jets. As a consequence, the $b_1b_2b_3b_4$ system acquires at NLO in QCD very large transverse momenta which are highly suppressed at LO.

\begin{figure}[t!]
        \centering        
        \includegraphics[width=0.48\linewidth]{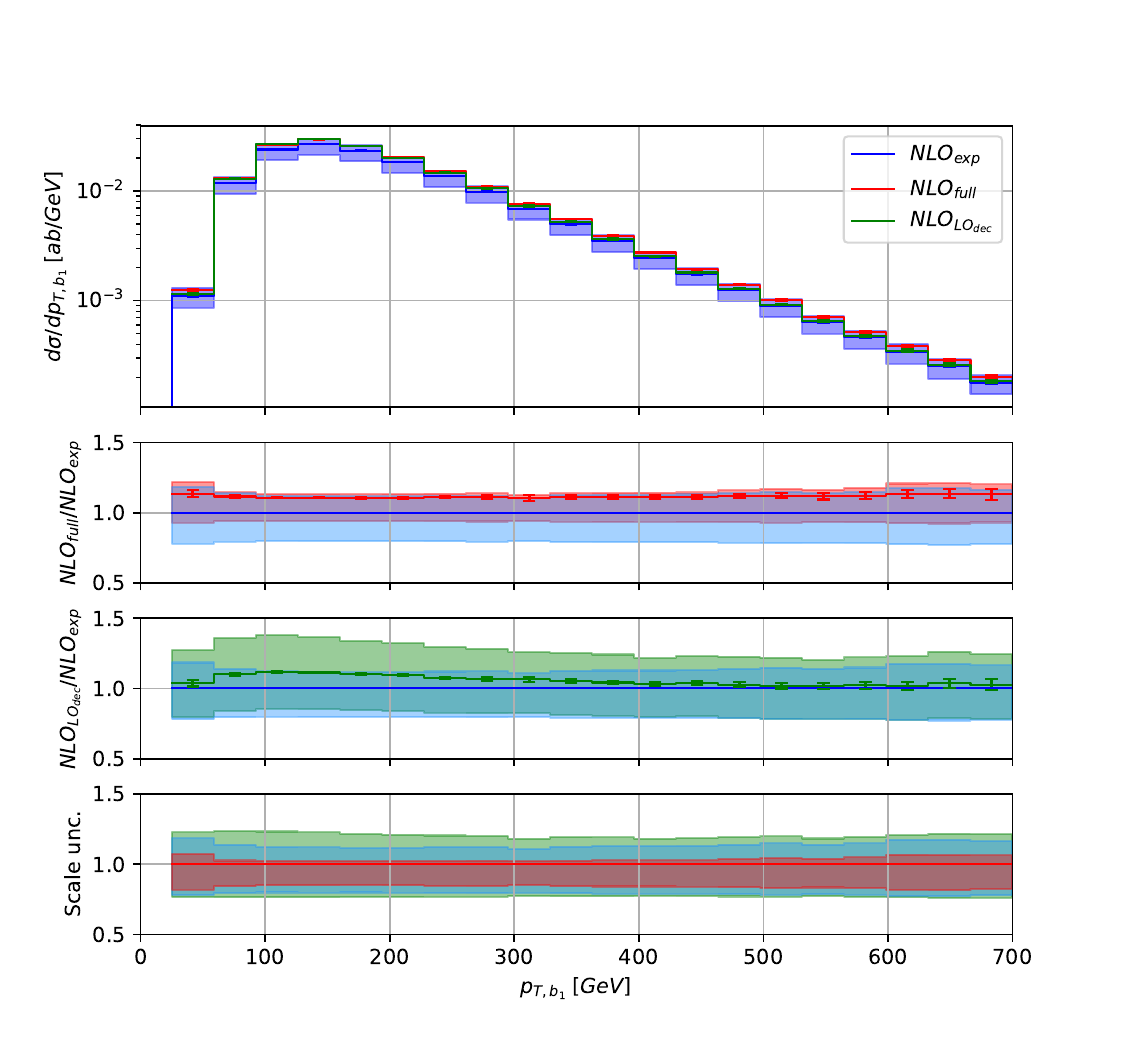}
        \includegraphics[width=0.48\linewidth]{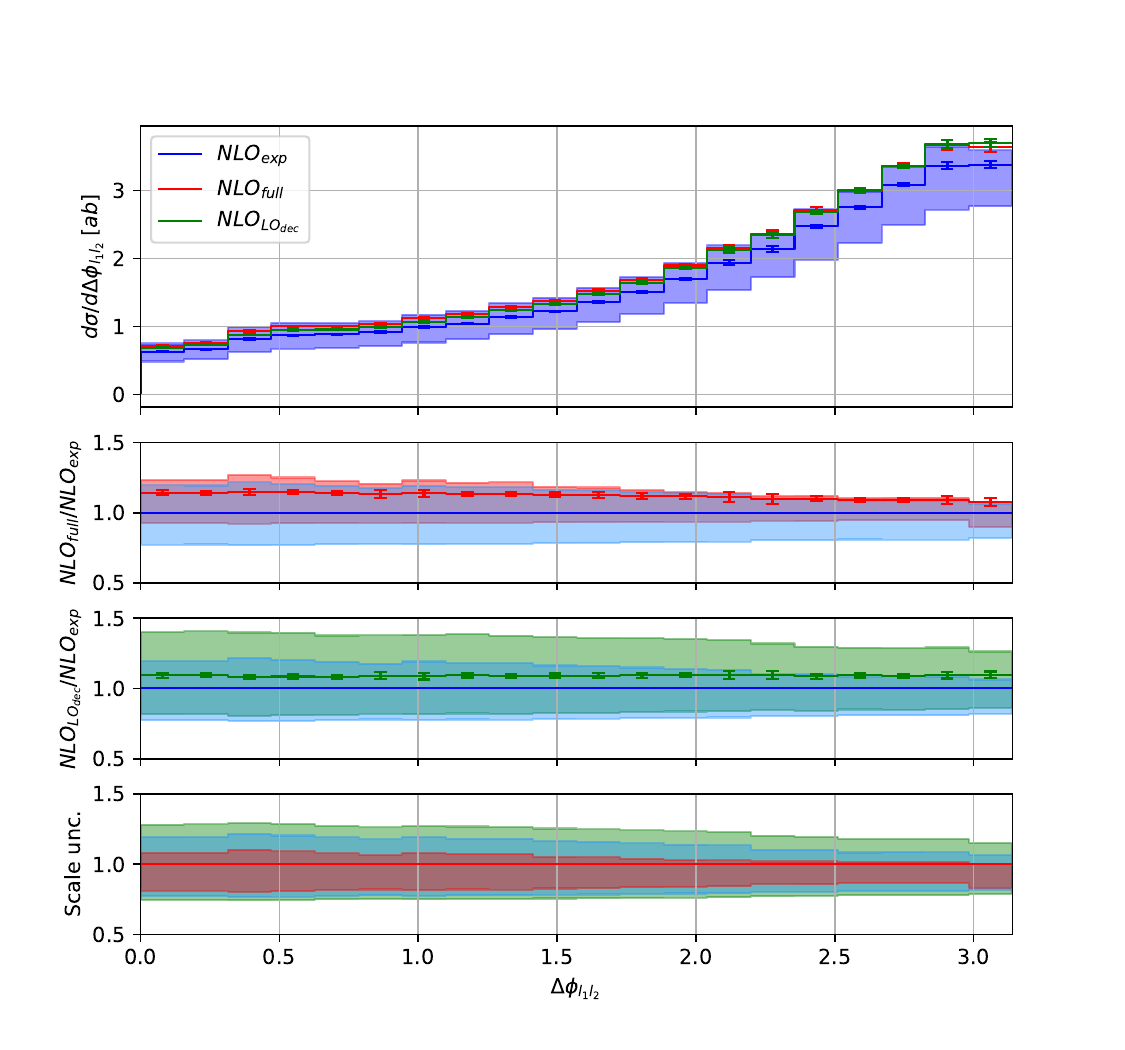}
        \caption{\textit{Differential cross-section distributions for the $pp \to t\bar{t}t\bar{t}$ process in the $4\ell$ channel for the $\rm NLO_{exp}$ (blue), $\rm NLO_{full}$ (red) and $\rm NLO_{LO_{dec}}$ (green) for the $p_{T, \, b_1}$ (left) and $\Delta \phi_{l_1l_2}$ (right) observables. The upper panels display the absolute predictions while in the middle panels the ratio to the expanded NWA is also provided, including the relative NLO scale uncertainties. In the bottom panels, the NLO scale uncertainties for the three approaches, normalised to their corresponding NLO results are also depicted. Figures are taken from Ref. \cite{Dimitrakopoulos:2024qib}.}}
         \label{nwa_methods.png}
\end{figure}

In Figure \ref{nwa_methods.png} we provide a differential comparison of the various NWA methods by displaying the $p_{T,\,b_1}$ and $\Delta \phi_{l_1l_2}$ observables. From the middle panels we observe that, for both cases, the impact of NLO QCD corrections on the decays is of the order of $10\%$. Similarly, the higher-order effects that are relevant in the $\rm NLO_{full}$ approach are of comparable size and remain relatively constant across the entire plotted ranges. Most importantly, the bottom panels reveal that including NLO QCD corrections in both the production and decays of the four top quarks leads to a significant reduction in the size of the NLO scale uncertainties. This reduction is observed across all regions of the phase space.

\section{Summary}
In this contribution, we have presented NLO QCD corrections for the four-top quark production process in the $4\ell$ channel using the NWA framework. We have found that neglecting NLO QCD corrections in the decays of the top quarks has overestimated the integrated fiducial NLO cross section by up to $9\%$. However, all various NWA treatments are compatible with each other as the results agree within their corresponding theoretical uncertainties. In addition, the inclusion of higher-order effects in the top-quark decays has decreased the size of the theoretical uncertainties both at the integrated and differential cross-section levels. At the differential cross-section level, the NLO QCD corrections are generally moderate for most of the observables we have examined. Nevertheless, there are some exceptions, such as $p_{T,\, b_1b_2b_3b_4}$, where the QCD corrections are quite large in specific regions of the phase space. To conclude, we would like to note that to assess the impact of hard emissions in the top-quark decays and the significance of NLO spin correlations, it is essential to compare our results to the NLO QCD results matched to PS. In future studies, we plan to perform such a comparison for this decay channel.

\section{Acknowledgments}
{This research was supported by the Deutsche Forschungsgemeinschaft (DFG) under grant 400140256 - GRK 2497: \textit{The physics of the heaviest particles at the LHC.}}
%
% \begin{thebibliography}{99}
% \bibitem{...}
% ....

% \end{thebibliography}
\bibliographystyle{JHEP}
\bibliography{references}

\end{document}